\documentclass[review, authoryear]{elsarticle}
\usepackage{array}
\usepackage{multirow}
\usepackage{amsmath}

\begin{document}

\begin{frontmatter}
\title{Copas' method is sensitive to different mechanisms of publication bias}
\author[1]{Osama Almalik}
\ead{o.almalik@tue.nl}

\author[1]{Zhuozhao Zhan}
\ead{z.zhan@tue.nl}

\author[1,2]{Edwin R. van den Heuvel\corref{cor1}%
\fnref{fn1}}
\ead{e.r.v.d.heuvel@tue.nl}

\cortext[cor1]{Corresponding author}
\fntext[fn1]{Address: Den Dolech 2, 5612 AZ Eindhoven, The Netherlands}
\address[1]{Department of Mathematics and Computer Science, Eindhoven University of Technology, Eindhoven, The Netherlands}
\address[2]{Department of Preventive Medicine and Epidemiology, School of Medicine, Boston University, Bost, USA}

\begin{abstract}
Copas' method corrects a pooled estimate from an aggregated
data meta-analysis for publication bias. Its performance has been
studied for one particular mechanism of publication bias. We show
through simulations that Copas' method is not robust against other
realistic mechanisms. This questions the usefulness of Copas' method,
since publication bias mechanisms are typically unknown in practice.
\end{abstract}
\end{frontmatter}

\section{Introduction}

In an aggregated data (AD) meta-analysis, published effect sizes from
similar research studies are collected to determine a precise pooled
effect size. When not all executed research studies are published,
an AD meta-analysis may lead to a biased estimate. To correct the
pooled estimate for this publication bias, various methods have been
proposed \citep{Jin2015, Mueller2016, Rucker2011}. Selection model approaches
implement a conditional or weighted likelihood function for estimation,
where the weights are based on the selection mechanism \citep{Hedges2006}.
Copas' selection method \citep{Copas2000,Copas2001} uses the standard
errors of the study effect sizes to create these weights. The method
gives a higher weight to studies with a lower probability of being
published.

Copas' method has been compared to the Trim and Fill method \citep{Duval2000,Duval2000a}
using 157 meta-analyses. Even though both methods produced similar
point estimates, Copas' method was preferred since it produced larger
standard errors, making Copas' method somewhat more conservative \citep{Schwarzer2010}.
Since direct likelihood-based methods may sometimes suffer from convergence
issues, an expectation-maximization (EM) algorithm was developed for
Copas' method \citep{Ning2017}. Furthermore, a Bayesian extension of
Copas' method was developed for network meta-analysis \citep{Mavridis2013}.
This all shows the importance of Copas' method in meta-analysis.

Unfortunately, the performance of Copas' method has been investigated
for one particular mechanism of publication bias using simulation
studies, even though other mechanisms for publication bias have been
proposed in literature \citep{Stanley2008,Stanley2014,Aert2018,Hedges1984,McShane2016}.
We will demonstrate that Copas' method is sensitive to these mechanisms
when mean differences are being pooled.

Section 2 describes Copas' method and three mechanisms for publication
bias. Section 3 describes our simulation study. The results and the
discussion are presented in Sections 4 and 5, respectively.

\section{Statistical methods}

The information in an AD meta-analysis consists of the pair $(D_{i},S_{i})$
for study $i=1,2,...,m$, where $D_{i}$ is the observed or collected
effect size and $S_{i}$ is the accompanied standard error. In some
applications there may also exist a degrees of freedom for the standard
error \citep{Cochran1954}, but this is ignored here.

\subsection{The Copas method}

\citet{Copas2000,Copas2001} considered a population of study
effect sizes that follow the random effects meta-analysis model
\begin{equation}
D_{i}=\theta+U_{i}+\varepsilon_{i},\label{eq:RE-model}
\end{equation}
with $\theta$ the unknown mean effect size of interest, $U_{i}\sim N(0,\tau^{2})$
the heterogeneity in study effect sizes, and $\varepsilon_{i}\sim N(0,\sigma_{i}^{2})$
the residual independent of $U_{i}$ with an unknown variance $\sigma_{i}^{2}$
that may vary with study. However, they assumed that only a selective
subset of all studies has been published and introduce a selection
model $Z_{i}=\alpha+\beta S_{i}^{-1}+\delta_{i}$, with $\alpha$
and $\beta$ fixed parameters, $\delta_{i}\sim N(0,1)$ being correlated
with $\varepsilon_{i}$, $\rho=\mathsf{CORR}(\varepsilon_{i},\delta_{i})$,
and $D_{i}$ only being published when $Z_{i}>0$. Note that studies
with smaller standard errors have a higher probability of being published
and when $\beta=0$ and $\alpha$ is large, there is no publication
bias present.

Based on the population and selection model for effect sizes, a weighted
or conditional log likelihood function is constructed
\[
\ell\left(\theta,\tau^{2},\rho\right)=\sum_{i=1}^{m}\left[\log p\left(D_{i}|Z_{i}>0,S_{i}\right)\right],
\]
with $p\left(D_{i}|Z_{i}>0,S_{i}\right)$ the conditional probability
density of an effect size given that the study is selected. Using
a joint normality assumption on $(\varepsilon_{i},\delta_{i})$ and
assuming that $(\varepsilon_{i},\delta_{i})$ is independent of $U_{i}$,
the conditional log likelihood function can be written in the following
explicit expression \citep{Copas2000,Copas2001}
\begin{equation}
\sum_{i=1}^{m}\left[-\tfrac{1}{2}\log\left(\tau^{2}+\sigma_{i}^{2}\right)-\dfrac{(D_{i}-\theta)^{2}}{2(\tau^{2}+\sigma_{i}^{2})}-\log\Phi(\alpha+\beta S_{i}^{-1})+\log\Phi(V_{i})\right],\label{eq:log-like}
\end{equation}
with $\Phi$ the standard normal distribution function, $V_{i}$ given
by $V_{i}=[\alpha+\beta S_{i}^{-1}+\tilde{\rho}_{i}(D_{i}-\theta)/(\tau^{2}+\sigma_{i}^{2})^{1/2}]/[1-\tilde{\rho}_{i}^{2}]^{1/2}$,
and $\tilde{\rho}_{i}=\sigma_{i}\rho/[\tau^{2}+\sigma_{i}^{2}]^{1/2}$.
The unknown variance $\sigma_{i}^{2}$ in \eqref{eq:log-like} is
replaced by $S_{i}^{2}/[1-c_{i}^{2}\rho^{2}]$, with $c_{i}=\lambda(\alpha+\beta S_{i}^{-1})[\alpha+\beta S_{i}^{-1}+\lambda(\alpha+\beta S_{i}^{-1})]$,
$\lambda(z)=\phi(z)/\Phi(z)$, and $\phi$ the standard normal density
function.

For fixed values of $\alpha$ and $\beta$, the log likelihood function
in \eqref{eq:log-like} is maximized over $\theta$, $\tau^{2}$,
and $\rho$ and their confidence intervals are based on asymptotic
theory. By studying a grid of different values for $\alpha$ and $\beta>0$,
such that $0.01\leq P(Z_{i}>0|S_{i})\leq0.99$ for the smallest and
largest value of $S_{i}$, the sensitivity of the pooled estimator
$\hat{\theta}$ on $\alpha$ and $\beta$ can be investigated \citep{Copas2000,Copas2001}.
Settings for $\alpha$ and $\beta$ for which selection bias is not
rejected would fit best with the data. This selection bias is tested
with a form of Egger's test \citep{Egger1997}. The random effects model
is extended to $D_{i}=\theta+\gamma S_{i}^{-1}+U_{i}+\varepsilon_{i}$
and $H_{0}:\gamma=0$ is tested with a likelihood ratio test \cite{Copas2000,Copas2001,Carpenter2009}.
We used the R-package ``copas'' which is part of the R-package meta
to carry out the Copas method \cite{Carpenter2009}.

\subsection{Selection models}

The selection model of Copas is based on the positiveness of the latent
variable $Z_{i}=\alpha+\beta S_{i}^{-1}+\delta_{i}$, with $\delta_{i}$
correlated with the residual in the random effect model in \eqref{eq:RE-model}.
However, there may be alternative approaches that would be based on
the standardized effect sizes $D_{i}/S_{i}$. Indeed, standardized
effect sizes closer to zero would be less likely to be published and
large effect sizes (at one side or in one direction) would be more
likely to be published \citep{Hedges1984}.

\subsubsection{Significant effect size}

Selection models based on the $p$-value of the study effect have
been proposed in literature \citep{Stanley2008,Stanley2014,Aert2018,Hedges1984,McShane2016}.
When the effect size is significant (assuming more positive effect
sizes), i.e., $D_{i}/S_{i}>z_{1-\alpha}$, with $\alpha$ the significance
level and $z_{q}$ the $q$\textsuperscript{th} quantile of a standard
normal distribution, the study is included. To add randomness to the
non-significant studies, a uniform distributed random variable $U(0,1)$
and a parameter $\pi_{\mathrm{pub}}$ can be used. If the uniform
random variable is smaller than or equal to $1-\pi_{\mathrm{pub}}$,
the non-significant study is included too, and otherwise it is excluded.

\subsubsection{Standardized effect size}

An alternative approach, is to use $D_{i}/S_{i}$ in a selection model
similar to Copas' selection model. Study $i$ is published when the
latent variable $Z_{i}=a+bD_{i}/S_{i}+\delta_{i}$ is positive, with
$a$ and $b$ fixed parameters, and with $\delta_{i}\sim N\left(0,1\right)$,
now being independent of the residual in model \eqref{eq:RE-model}.
We do not need a non-zero correlation between $\delta_{i}$ and $\varepsilon_{i}$,
since the correlation with the population effect size or the selection
of studies is now directly induced by the standardized effect size.
The probability that study $i$ is selected is $P(Z_{i}>0|D_{i}=d,S_{i}=s)=\Phi(a+bd/s)$.

\subsection{Simulation model}

We will first draw a population of effect sizes and standard errors,
i.e., draw pair $(D_{i},S_{i})$, that is calculated from individual
participant data (IPD) for two groups in each study. Then we will
use the different selection models to eliminate studies from the population.

\subsubsection{Population of aggregated data}

We consider a meta-analysis with $m$ studies, having sample sizes
$n_{i}$, $i=1,\cdots,m$. The number of participants $n_{i}$ for
study $i$ is drawn using an overdispersed Poisson distribution with
parameter $\lambda$. The value $\gamma_{i}\sim\Gamma\left(a_{0},b_{0}\right)$,
with $\Gamma\left(a_{0},b_{0}\right)$ a gamma distribution with parameters
$a_{0}$ and $b_{0}$, is drawn to make a study specific parameter
$\lambda_{i}=\lambda\exp\left(0.5\gamma_{i}\right)$. Then $n_{i}$
is drawn from a Poisson distribution with parameter $\lambda_{i}$,
i.e., $n_{i}\sim\mathrm{Pois}\left(\lambda_{i}\right)$. This sample
size is then split in two sample sizes using a Binomial distribution
with parameter $p$, i.e., $n_{i0}\sim\mathrm{Bin}(n_{i},p)$ and
$n_{i1}=n_{i}-n_{i0}$.

Then a continuous response $Y_{ijk}$ for individual $k(=1,\cdots,n_{ij})$,
in group $j(=0,1)$, for study $i(=1,2,...,m)$ is simulated according
to a linear mixed model:
\begin{equation}
Y_{ijk}=\mu+\beta_{j}+U_{ij}+\epsilon_{ijk},\label{eq:general model-1}
\end{equation}
with $\mu$ a general mean, $\beta_{j}$ an effect of group $j$ ($\beta_{0}=0$
and $\beta_{1}=\theta$), $U_{ij}$ a study-specific random effect
for group $j$, and residual $\epsilon_{ijk}\sim N\left(0,\zeta^{2}\right)$.
We assume that $(U_{i0},U_{i1})^{T}$ is bivariate normally distributed
with zero means and variance-covariance matrix $\varSigma$ given
by
\[
\varSigma=\left[\begin{array}{cc}
\sigma_{0}^{2} & \rho_{01}\sigma_{0}\sigma_{1}\\
\rho_{01}\sigma_{0}\sigma_{1} & \sigma_{1}^{2}
\end{array}\right].
\]

After simulating the individual responses, the study effect size is
calculated by the mean difference $D_{i}=\bar{Y}_{i0.}-\bar{Y}_{i1.}$,
with $\bar{Y}_{ij.}=\sum_{i=1}^{n_{ij}}Y_{ijk}/n_{ij}$ the average
of group $j$ in study $i$. It is straightforward to see that $D_{i}$
satisfies model \eqref{eq:RE-model} with $U_{i}=U_{i0}-U_{i1}\sim N(0,\sigma_{0}^{2}-2\rho_{01}\sigma_{0}\sigma_{1}+\sigma_{1}^{2})$
and $\varepsilon_{i}\sim N(0,\zeta^{2}[n_{i0}^{-1}+n_{i1}^{-1}])$.
The standard error $S_{i}$ was estimated using the formula $S_{i}=\sqrt{S_{i0}^{2}/n_{i0}+S_{i1}^{2}/n_{i1}}$,
with $S_{ij}^{2}=\sum_{k=1}^{n_{ij}}(Y_{ijk}-\bar{Y}_{ij.})^{2}/(n_{ij}-1)$
the sample variance of group $j$ in study $i$, not assuming that
the residual variance in model \eqref{eq:general model-1} is homogeneous.

The settings of the parameters are chosen such that the simulation
corresponds approximately with a meta-analysis of clinical trials
on hypertension treatment. Parameter settings used to generate the
aggregated data are $m\in\{30,50,100\}$, $\lambda=100$, $a_{0}=b_{0}=1$,
$p=0.5$, $\mu=160$, $\theta=-0.5$, $\zeta^{2}=100$, $\sigma_{0}^{2}\in\{0,2\}$,
$\sigma_{1}^{2}\in\{0,3\}$, and $\rho_{01}\in\{0,0.7\}$. We will
run all combinations of parameter choices and simulate 1000 meta-analysis
studies. Note that this implies that we study five levels of heterogeneity,
i.e., $\tau^{2}=\sigma_{0}^{2}-2\rho_{01}\sigma_{0}\sigma_{1}+\sigma_{1}^{2}\in\{0,2,5-1.4\sqrt{6},3,5\}$,
but we will only report three levels $\{0,5-1.4\sqrt{6},5\}$. These
settings correspond to an intraclass correlation coefficient (ICC)
of approximately 0\%, 40\%, and 68\%, respectively, since we expect
an average sample size per treatment group to be equal to $85$ individuals.

\subsubsection{Selection of studies}

Copas' selection model requires simulation of $Z_{i}=\alpha+\beta S_{i}^{-1}+\delta_{i}$,
with $\delta_{i}$ being correlated to $\varepsilon_{i}$ in \eqref{eq:RE-model}.
The residual $\varepsilon_{i}$ can be calculated from the simulation
of the individual data, since $\varepsilon_{i}=\bar{\epsilon}_{i0.}-\bar{\epsilon}_{i1.}$,
with $\bar{\epsilon}_{ij}=\sum_{k=1}^{n_{ij}}\epsilon_{ijk}/n_{ij}$.
Then $\delta_{i}$ can be drawn from a normal distribution
\[
\delta_{i}|\bar{\epsilon}_{i0.}-\bar{\epsilon}_{i1.}\sim N\left(\rho[\bar{\epsilon}_{i0.}-\bar{\epsilon}_{i1.}]/\sqrt{\zeta^{2}[n_{i0}^{-1}+n_{i1}^{-1}]},1-\rho^{2}\right),
\]
where $\rho=\mathsf{CORR}(\delta_{i},\varepsilon_{i})$ is the correlation
parameter taken equal to $\rho\in\{0,0.9\}$. The parameters $\alpha$
and $\beta$ will depend on the simulated population data and vary
with each simulation run.

We used the 5\% and 95\% quantiles of the set of precision estimates
$S_{1}^{-1}$, $S_{2}^{-1}$, ..., $S_{m}^{-1}$ for one meta-analysis,
say $q_{5}$ and $q_{95}$, respectively. The values $\alpha$ and
$\beta$ are chosen such that $P(Z_{i}>0|S_{i}^{-1}=q_{95})=0.99$
and $P(Z_{i}>0|S_{i}^{-1}=q_{5})=p_{0}$, with $p_{0}\leq0.50$. A
study with a small standard error is almost always selected, while
studies with larger standard errors are more likely eliminated from
the meta-analysis. Solving the two equations results in parameters
$\alpha\approx(z_{p_{0}}q_{95}-2.33q_{5})/(q_{95}-q_{5})$ and $\beta\approx(z_{p_{0}}-\alpha)/q_{5}$,
when the random term $\delta_{i}$ is independent of all other terms.
A study $i$ was selected if $Z_{i}>0$, and it was eliminated when
$Z_{i}\leq0$. We tuned the parameter $p_{0}$ such that we select
approximately 70\% of all simulated studies under the same settings.

Simulation of the selection models based on standardized effect sizes
$D_{i}/S_{i}$ are more straightforward. For latent variable $Z_{i}=a+bD_{i}/S_{i}+\delta_{i}$,
we draw $\delta_{i}$ from a standard normal distribution, independent
of anything else. Here we use $a$ and $b$ in the same way as $\alpha$
and $\beta$, but the quantiles $q_{5}$ and $q_{95}$ are now calculated
from the set of standardized effect sizes $D_{1}/S_{1}$, $D_{2}/S_{2}$,...,
$D_{m}/S_{m}$ (assuming $D_{i}$' s are mostly positive, otherwise
we could use $-D_{i}/S_{i}$). For the $p$-value based selection
of studies, we searched for values of $\pi_{\mathrm{pub}}$ such that
approximately 70\% of the studies are included.

The average effective number of studies $\bar{m}$ included in the
simulations for the three selection models will be reported.

\section{Results}

\begin{figure}[h]
\centering

\includegraphics[scale=0.33]{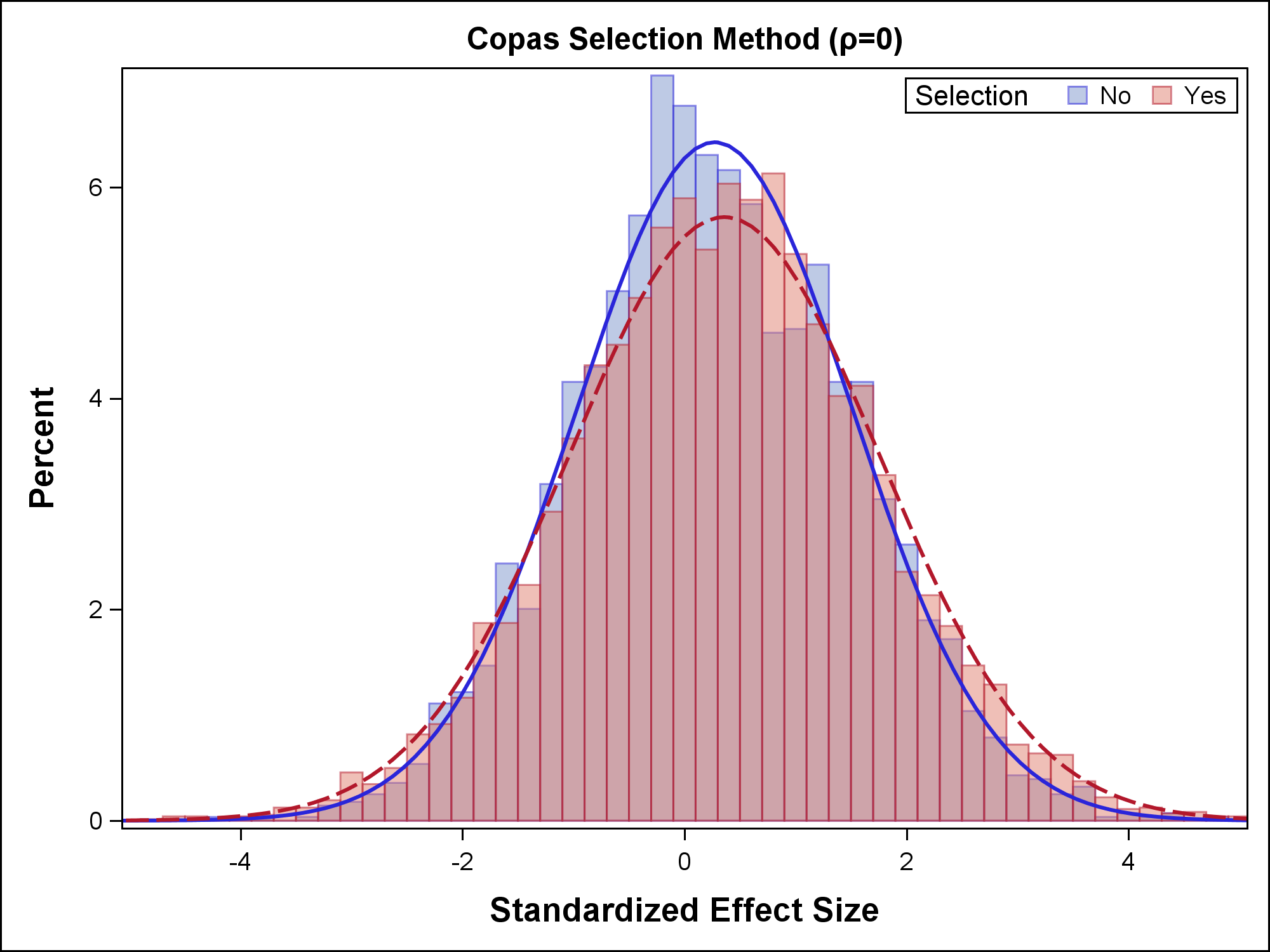}\includegraphics[scale=0.33]{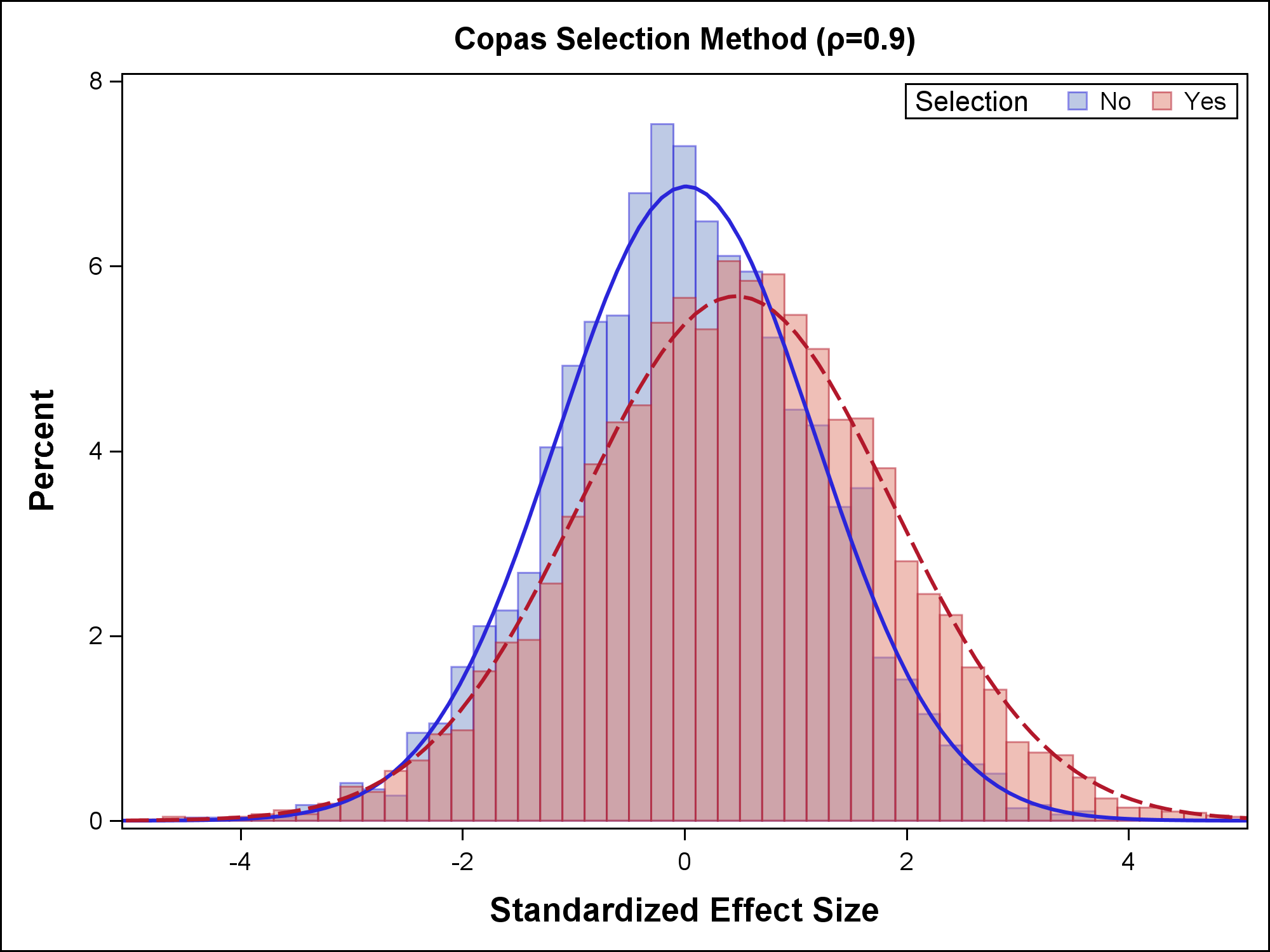}

\includegraphics[scale=0.33]{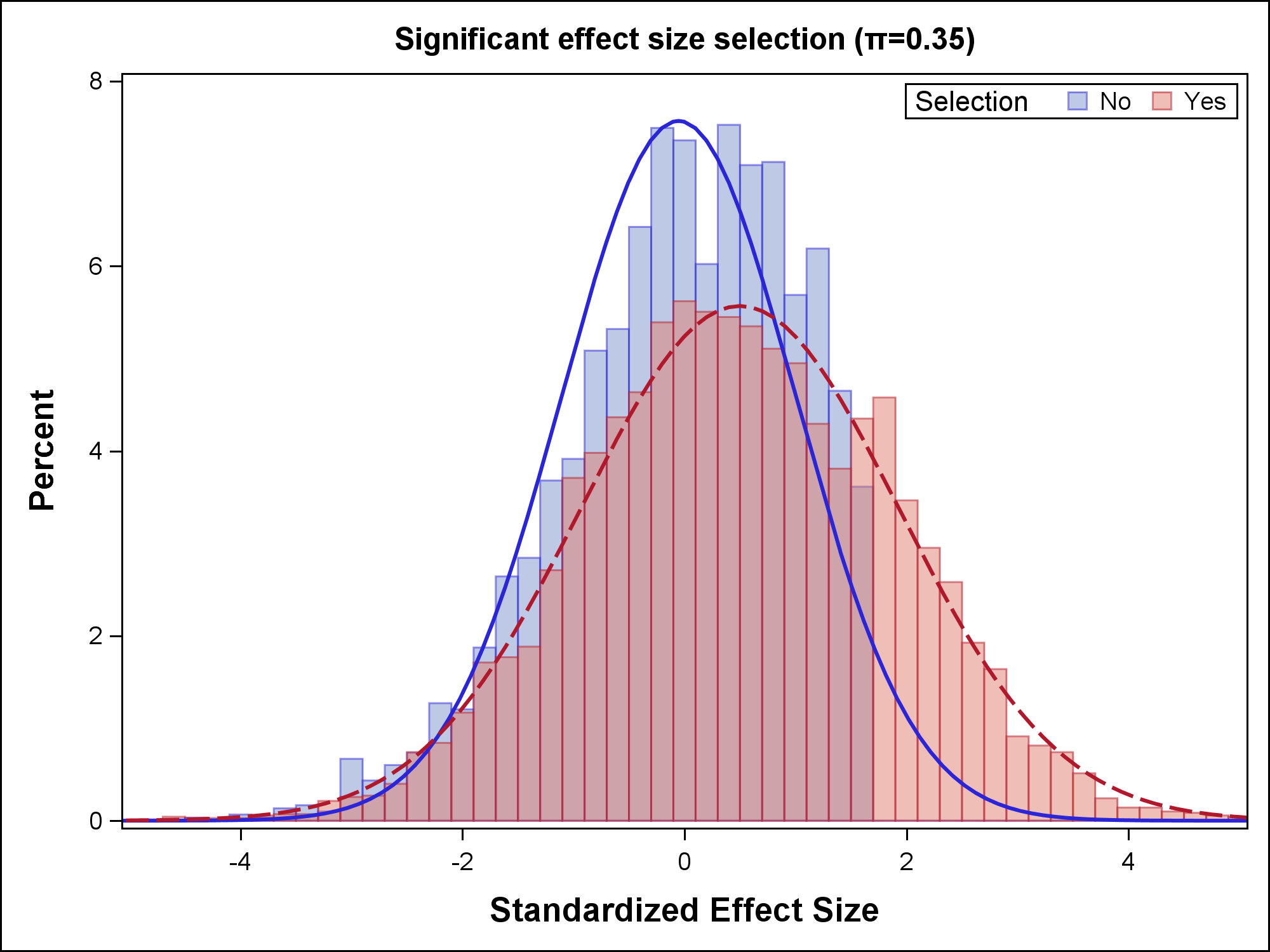}\includegraphics[scale=0.33]{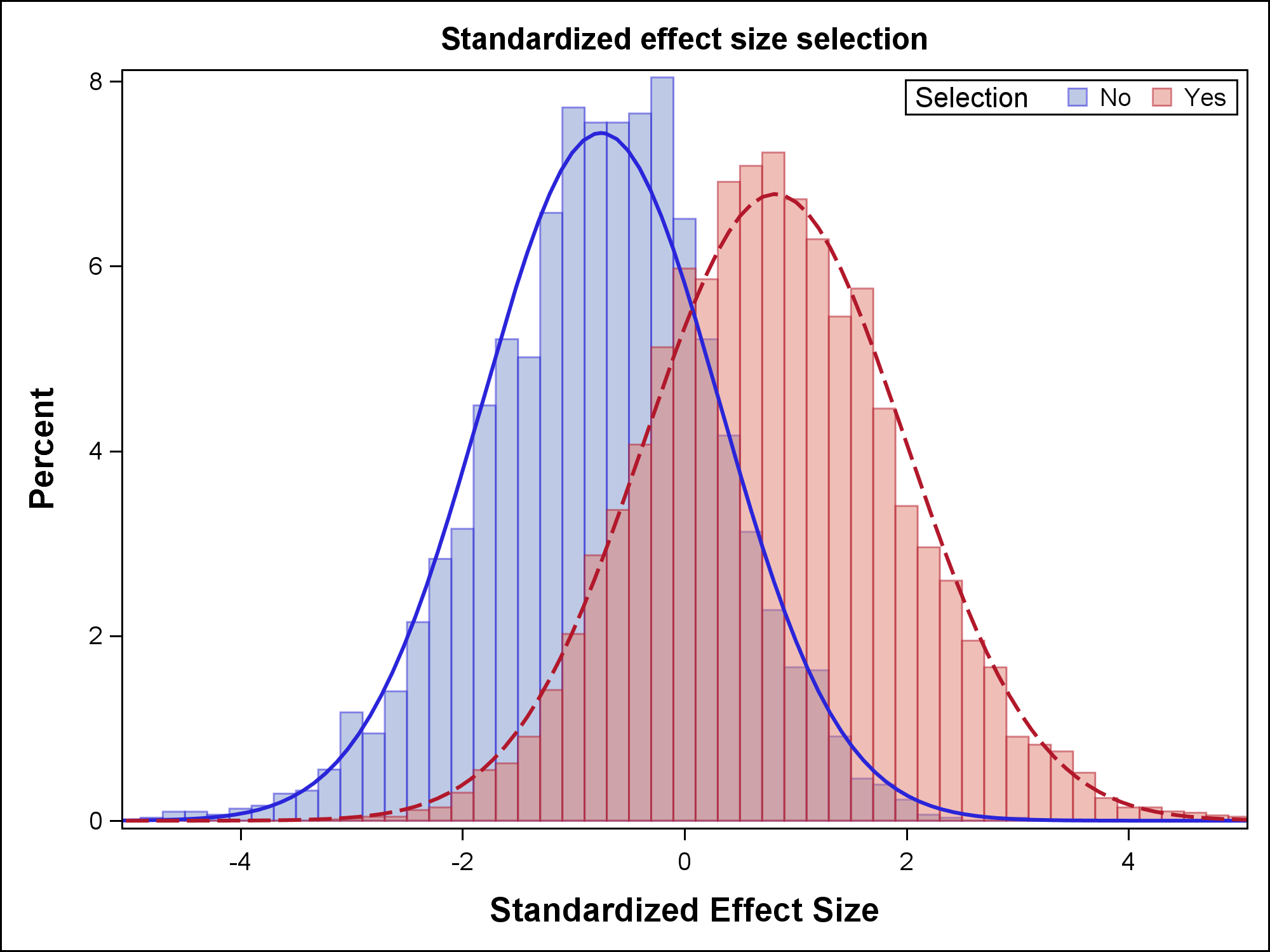}

\caption{Visualization of the selection of studies for the selection models.\label{fig:Visualization}}
\end{figure}

Figure \ref{fig:Visualization} shows the distributions of the standardized
effect sizes for the selected and non-selected studies for the four
selections models ($\sigma_{0}^{2}=2$; $\sigma_{1}^{2}=3$; $\rho_{01}=0$).
The mechanisms based on the standardized effect sizes have a stronger
effect on selection of studies than Copas' selection model. The selection
model based on $D_{i}/S_{i}$ also show a different mechanism. The
$p$-value based selection model shows the truncation of being significant,
$Z_{i}=a+bD_{i}/S_{i}+\delta_{i}$ shifts the distribution, while
Copas' selection models essentially eliminate higher standardized
effects sizes with lower probabilities.

The performance of Copas' method on estimation of the pooled effect
size ($\theta$) for the different selection methods is evaluated
with the Mean Squared Error (MSE), the bias, and the coverage probability
(CP). The results of the simulations are presented in Table \ref{tab:Performance-of-Copas}
for $m=30$. The results for other numbers of study sizes are very
similar to the results of $m=30$.

\begin{table}[h]
\caption{Performance of Copas' method (MSE, bias and CP(\%)) for estimation
of the pooled estimate ($\theta=-0.5$; publication rate $\approx70\%$;
$m=30$).\label{tab:Performance-of-Copas}}

\centering

\begin{tabular}{|c|c|c|l|l|c|c|c|c|}
\hline 
$\sigma_{0}^{2}$ & $\sigma_{1}^{2}$ & $\rho_{01}$ & \multicolumn{2}{l|}{Selection method} & MSE & Bias & CP(\%) & $\bar{m}$\tabularnewline
\hline 
\hline 
0 & 0 & 0 & \multirow{2}{*}{Copas} & $\rho=0$ & 0.11645 & 0.00393 & 0.943 & 20.7\tabularnewline
\cline{1-3} \cline{2-3} \cline{3-3} \cline{5-9} \cline{6-9} \cline{7-9} \cline{8-9} \cline{9-9} 
0 & 0 & 0 &  & $\rho=0.9$ & 0.14352 & -0.04650 & 0.922 & 21.1\tabularnewline
\hline 
0 & 0 & 0 & \multicolumn{2}{l|}{Significant effect} & 0.15348 & -0.03456 & 0.924 & 21.6\tabularnewline
\hline 
0 & 0 & 0 & \multicolumn{2}{l|}{Standardized effect} & 0.33874 & -0.45506 & 0.681 & 20.7\tabularnewline
\hline 
\hline 
2 & 3 & 0.7 & \multirow{2}{*}{Copas} & $\rho=0$ & 0.28760 & -0.01156 & 0.873 & 20.7\tabularnewline
\cline{1-3} \cline{2-3} \cline{3-3} \cline{5-9} \cline{6-9} \cline{7-9} \cline{8-9} \cline{9-9} 
2 & 3 & 0.7 &  & $\rho=0.9$ & 0.31806 & -0.05903 & 0.858 & 21.1\tabularnewline
\hline 
2 & 3 & 0.7 & \multicolumn{2}{l|}{Significant effect} & 0.39669 & -0.11541 & 0.875 & 21.0\tabularnewline
\hline 
2 & 3 & 0.7 & \multicolumn{2}{l|}{Standardized effect} & 0.77575 & -0.69885 & 0.521 & 20.6\tabularnewline
\hline 
\hline 
2 & 3 & 0 & \multirow{2}{*}{Copas} & $\rho=0$ & 0.50614 & -0.00193 & 0.886 & 20.7\tabularnewline
\cline{1-3} \cline{2-3} \cline{3-3} \cline{5-9} \cline{6-9} \cline{7-9} \cline{8-9} \cline{9-9} 
2 & 3 & 0 &  & $\rho=0.9$ & 0.55830 & -0.11847 & 0.877 & 21.1\tabularnewline
\hline 
2 & 3 & 0 & \multicolumn{2}{l|}{Significant effect} & 0.67515 & -0.14357 & 0.885 & 21.3\tabularnewline
\hline 
2 & 3 & 0 & \multicolumn{2}{l|}{Standardized effect} & 1.59896 & -1.03069 & 0.480 & 20.5\tabularnewline
\hline 
\end{tabular}
\end{table}

Introducing publication bias according to Copas' selection model,
clearly results in the lowest MSE and bias (as expected). When heterogeneity
in study effect sizes increases and when the selection model is correlated
to the random effects model ($\rho=0.9$) a bias appears that can
reach a level of 20\% of the pooled effect size. The coverage probability
is in general liberal and only close to nominal for homogeneous study
effect sizes. Selection of studies based on significant effect sizes
increases the MSE and the bias. For homogeneous study effect sizes,
the bias is still limited to approximately 7\%, but when heterogeneity
is increasing the relative bias can easily increase to approximately
30\%. Due to the increased MSE compared to Copas' selection model,
the coverage of the 95\% confidence interval remains at the same level
as Copas' selection model. When the selection is based directly on
the standardized effect sizes, Copas' model seem to fail completely,
in particular when heterogeneity is present. Copas' method does not
correct the estimate enough, leading to very high biases and low coverage
probabilities.

\section{Discussion}

The purpose of this paper was to investigate the performance of Copas'
method for adjusting the pooled estimate from an aggregated data meta
analysis in the presence of publication bias. We focused on effect
sizes in the form of mean differences and studied three different
selection models for publication bias. These selection models were
all (indirectly or directly) related to the effect size of a study
\citep{Hedges1984,McShane2016}.

Copas' method overestimates treatment effect (e.g., does not correct
enough) in case of between-study heterogeneity, regardless of the
selection model. The Copas method performs best and corrects adequately
when publication bias follows Copas' selection model. Our results
are comparable to results on bias and coverage in literature \citep{Ning2017}.
However, when the mechanism behind publication bias is different from
that used in the Copas' selection model, the method performs rather
poorly. This happens in particular when the standardized effect size
is the statistic that would drive publication bias. Heterogeneity
in study effect sizes emphasizes the shortcomings of Copas' method.

This paper only considered mean differences, but we do not think that
other types of effect sizes (e.g., log odds ratios) would provide
any different results. It is very common to assume that other types
of effect sizes also follow the random effects model in \eqref{eq:RE-model}
approximately, i.e., the model we used for our simulations. Additionally,
other reasons for publication bias, which we did not study, have been
mentioned in literature as well \citep{Sterne2011}, e.g., language bias,
availability bias, and cost bias. It is unknown how Copas' method
deals with these forms of biases, but we feel that it is unlikely
that Copas' method corrects appropriately, since these biases are
probably not described well by Copas' selection model. We recommend
to improve Copas' method to make it more robust against different
forms of publication bias.

\section*{Acknowledgments}

This research was funded by grant number 023.005.087 from the Netherlands
Organization for Scientific Research.

\section*{Conflict of Interest}
The authors have declared no conflict of interest.

\section*{Reference}
\bibliographystyle{elsarticle-harv}
\bibliography{ref}

\begin{thebibliography}{20}
\expandafter\ifx\csname natexlab\endcsname\relax\def\natexlab#1{#1}\fi
\expandafter\ifx\csname url\endcsname\relax
  \def\url#1{\texttt{#1}}\fi
\expandafter\ifx\csname urlprefix\endcsname\relax\def\urlprefix{URL }\fi

\bibitem[{Carpenter et~al.(2009)Carpenter, Rücker, and
  Schwarzer}]{Carpenter2009}
Carpenter, J., Rücker, G., Schwarzer, G., 2009. copas: An r package for
  fitting the copas selection model. The R Journal 1~(2), 31.

\bibitem[{Cochran(1954)}]{Cochran1954}
Cochran, W.~G., mar 1954. The combination of estimates from different
  experiments. Biometrics 10~(1), 101.

\bibitem[{Copas and Shi(2001)}]{Copas2001}
Copas, J., Shi, J., aug 2001. A sensitivity analysis for publication bias in
  systematic reviews. Statistical Methods in Medical Research 10~(4), 251--265.

\bibitem[{Copas and Shi(2000)}]{Copas2000}
Copas, J., Shi, J.~Q., sep 2000. Meta-analysis, funnel plots and sensitivity
  analysis. Biostatistics 1~(3), 247--262.

\bibitem[{Duval and Tweedie(2000{\natexlab{a}})}]{Duval2000a}
Duval, S., Tweedie, R., mar 2000{\natexlab{a}}. A nonparametric "trim and fill"
  method of accounting for publication bias in meta-analysis. Journal of the
  American Statistical Association 95~(449), 89.

\bibitem[{Duval and Tweedie(2000{\natexlab{b}})}]{Duval2000}
Duval, S., Tweedie, R., jun 2000{\natexlab{b}}. Trim and fill: A simple
  funnel-plot-based method of testing and adjusting for publication bias in
  meta-analysis. Biometrics 56~(2), 455--463.

\bibitem[{Egger et~al.(1997)Egger, Smith, Schneider, and Minder}]{Egger1997}
Egger, M., Smith, G.~D., Schneider, M., Minder, C., sep 1997. Bias in
  meta-analysis detected by a simple, graphical test. {BMJ} 315~(7109),
  629--634.

\bibitem[{Hedges(1984)}]{Hedges1984}
Hedges, L.~V., mar 1984. Estimation of effect size under nonrandom sampling:
  The effects of censoring studies yielding statistically insignificant mean
  differences. Journal of Educational Statistics 9~(1), 61--85.

\bibitem[{Hedges and Vevea(2006)}]{Hedges2006}
Hedges, L.~V., Vevea, J., mar 2006. Selection method approaches. In:
  Publication Bias in Meta-Analysis. John Wiley {\&} Sons, Ltd, pp. 145--174.

\bibitem[{Jin et~al.(2015)Jin, Zhou, and He}]{Jin2015}
Jin, Z.-C., Zhou, X.-H., He, J., nov 2015. Statistical methods for dealing with
  publication bias in meta-analysis. Statistics in Medicine 34~(2), 343--360.

\bibitem[{Mavridis et~al.(2013)Mavridis, Sutton, Cipriani, and
  Salanti}]{Mavridis2013}
Mavridis, D., Sutton, A., Cipriani, A., Salanti, G., jul 2013. A fully bayesian
  application of the copas selection model for publication bias extended to
  network meta-analysis. Statistics in Medicine 32~(1), 51--66.

\bibitem[{McShane et~al.(2016)McShane, Böckenholt, and Hansen}]{McShane2016}
McShane, B.~B., Böckenholt, U., Hansen, K.~T., sep 2016. Adjusting for
  publication bias in meta-analysis. Perspectives on Psychological Science
  11~(5), 730--749.

\bibitem[{Mueller et~al.(2016)Mueller, Meerpohl, Briel, Antes, von Elm, Lang,
  Motschall, Schwarzer, and Bassler}]{Mueller2016}
Mueller, K.~F., Meerpohl, J.~J., Briel, M., Antes, G., von Elm, E., Lang, B.,
  Motschall, E., Schwarzer, G., Bassler, D., dec 2016. Methods for detecting,
  quantifying, and adjusting for dissemination bias in meta-analysis are
  described. Journal of Clinical Epidemiology 80, 25--33.

\bibitem[{Ning et~al.(2017)Ning, Chen, and Piao}]{Ning2017}
Ning, J., Chen, Y., Piao, J., feb 2017. Maximum likelihood estimation and em
  algorithm of copas-like selection model for publication bias correction.
  Biostatistics 18~(3), 495--504.

\bibitem[{Rucker et~al.(2011)Rucker, Schwarzer, Carpenter, Binder, and
  Schumacher}]{Rucker2011}
Rucker, G., Schwarzer, G., Carpenter, J.~R., Binder, H., Schumacher, M., jul
  2011. Treatment-effect estimates adjusted for small-study effects via a limit
  meta-analysis. Biostatistics 12~(1), 122--142.

\bibitem[{Schwarzer et~al.(2010)Schwarzer, Carpenter, and
  Rücker}]{Schwarzer2010}
Schwarzer, G., Carpenter, J., Rücker, G., mar 2010. Empirical evaluation
  suggests copas selection model preferable to trim-and-fill method for
  selection bias in meta-analysis. Journal of Clinical Epidemiology 63~(3),
  282--288.

\bibitem[{Stanley(2008)}]{Stanley2008}
Stanley, T.~D., 2008. Meta-regression methods for detecting and estimating
  empirical effects in the presence of publication selection. Oxford Bulletin
  of Economics and Statistics 70~(1), 103--127.

\bibitem[{Stanley and Doucouliagos(2014)}]{Stanley2014}
Stanley, T.~D., Doucouliagos, H., sep 2014. Meta-regression approximations to
  reduce publication selection bias. Research Synthesis Methods 5~(1), 60--78.

\bibitem[{Sterne et~al.(2011)Sterne, Sutton, Ioannidis, Terrin, Jones, Lau,
  Carpenter, Rucker, Harbord, Schmid, Tetzlaff, Deeks, Peters, Macaskill,
  Schwarzer, Duval, Altman, Moher, and Higgins}]{Sterne2011}
Sterne, J. A.~C., Sutton, A.~J., Ioannidis, J. P.~A., Terrin, N., Jones, D.~R.,
  Lau, J., Carpenter, J., Rucker, G., Harbord, R.~M., Schmid, C.~H., Tetzlaff,
  J., Deeks, J.~J., Peters, J., Macaskill, P., Schwarzer, G., Duval, S.,
  Altman, D.~G., Moher, D., Higgins, J. P.~T., jul 2011. Recommendations for
  examining and interpreting funnel plot asymmetry in meta-analyses of
  randomised controlled trials. {BMJ} 343~(jul22 1), d4002--d4002.

\bibitem[{van Aert and van Assen(2018)}]{Aert2018}
van Aert, R. C.~M., van Assen, M. A. L.~M., oct 2018. Correcting for
  publication bias in a meta-analysis with the p-uniform* method.

\end{thebibliography}

\end{document}